# Logic Gates with Ion Transistors

H. Grebel

Electronic Imaging Center and the Electrical and Computer Engineering Department at NJIT, Newark, NJ 07102. grebel@njit.edu

**Abstract:** Electronic logic gates are the basic building blocks of every computing and micro controlling system. Logic gates are made of switches, such as diodes and transistors. Ion-selective, ionic switches may emulate electronic switches [1-8]. If we ever want to create artificial bio-chemical circuitry, then we need to move a step further towards ion-logic circuitry. Here we demonstrate ion XOR and OR gates with electrochemical cells, and specifically, with two wet-cell batteries. In parallel to vacuum tubes, the batteries were modified to include a third, permeable and conductive mid electrode (the gate), which was placed between the anode and cathode in order to affect the ion flow through it. The key is to control the cell output with a much smaller biasing power, as demonstrated here. A successful demonstration points to self-powered ion logic gates.

**Introduction:** Electrochemical reactions have been studied since the early eighteenth century [9-10]. Two half-cell reactions are considered. In one, oxidation of the anode takes place. Excess electrons then flow through an external load to the second half-cell, where reduction takes place at the cathode. The circuit is completed by ionic current in the electrolyte. The two-half cells are connected via a permeable membrane, which enables the passage of ions, yet, limits the flow of the bulk electrolyte molecules.

We wish to control the ion flow inside electrochemical cells, electrically. Control of a reaction near an electrode (working electrode) is routinely made with an auxiliary electrode and a saturated reference electrode using potentiostats or galvanostats. This approach may affect the surface potential of the working electrode and the control process could become nonlinear. Our approach is different: here, a third permeable electrode (the gate electrode) is placed between the anode and the cathode. Upon biasing of this mid-electrode we form an electrolyte barrier to the flowing ions. Consequently, the external current and voltage of the cell are controlled [11-13].

Our approach is also different than what is accustomed to in the literature. The latter are typically conducted with functionalized membranes [14-15] for the purpose of ion separation. A bipolar Ion Transistor reported in Ref. 1 is a good example – the design was based on ion-selective membranes, and hence was ion specific. In contrast, we aim at controlling both anions and cations by the same electrolyte barrier potential.

**Simulations:** Simulations employed a commercial tool, based on finite elements (COMSOL). We used a very simple Zn-Pt cell: a Zn electrode as the anode and a Pt electrode as the cathode. The model allowed us to deal with a single ion component ($Zn^{2+}$) and took into account the reactions at the anode (oxidation of Zn) and on the cathode (formation of hydrogen), yet assumed no reaction at the gate. The diffusion of ions in the cell has considered only excess $Zn^{2+}$ ions in the electrolyte. The local ion current density was assessed as the negative spatial derivative of the local electrolyte potential (which is proportional to the local electric field) multiplied by the electrolyte conductivity. The effective electrolyte-to-metallic volume ratio in the porous electrode was 1:1. Other simulation parameters were: electrical conductivity of Pt, Zn, porous electrode and $Zn^{2+}$ concentration in the electrolyte, respectively: $10^8$, $10^7$, $3\times10^5$, 0.01 S/m. The upper tip of the Pt cathode was grounded and the upper tip of the Zn anode was kept at (-0.8) V, slightly lower than the standard potential of the Zn anode ($E_0^{(Zn)}$=-0.82 V). This means that the cell's voltage (between Pt cathode and the Zn anode) was +0.8 V. Results are shown in Figs. 1, 2. In the absence of gate reaction, the gate voltage that stops the battery from functioning (the stopping potential) is 0 V when the gate is biased with respect to the grounded cathode. The stopping potential is -2 V when the gate is biased with respect to the grounded anode. The external electrical cell current is negative (meaning flowing towards the anode) and its slope is negative, similar to Fig. 1a. We may conclude that: (1) changes in the electrolyte potential at the gate affect the external current. (2) There are gate bias conditions which can effectively stop the external cell's current from flowing.

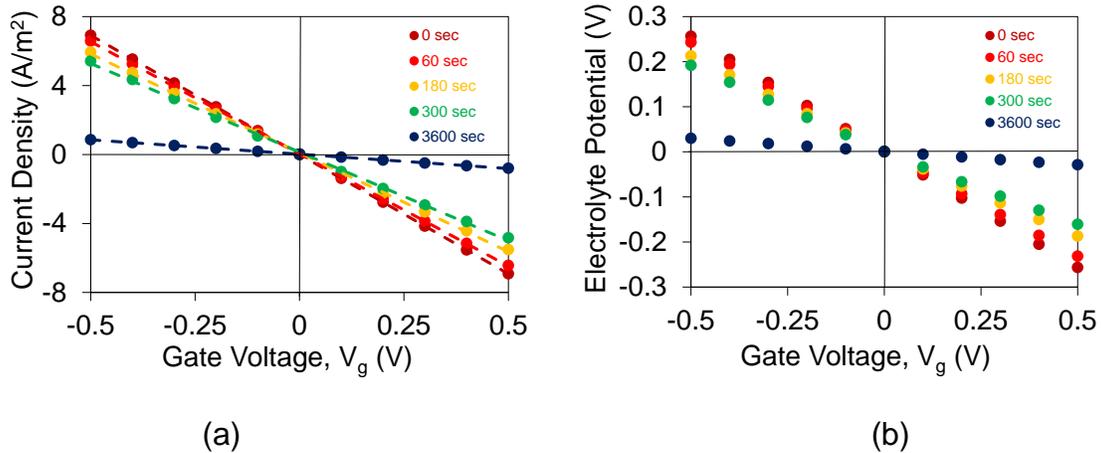

(a)             (b)

Fig. 1: (a) External cell's current density as a function of gate voltage. The current density is presented at various times. (b) Electrolyte potential at the mid-gate position as a function the gate voltage. The electrolyte potential is presented at various times.

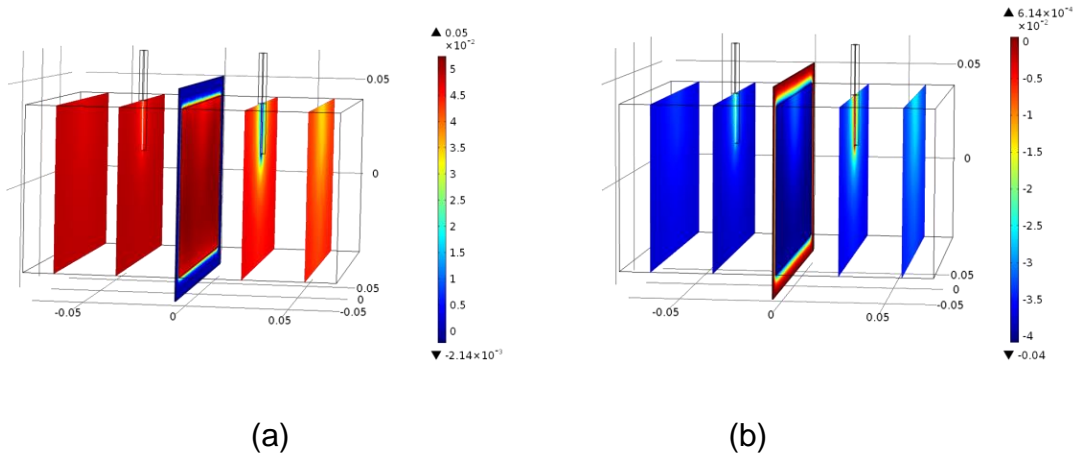

(a)                          (b)

Fig. 2: Slices of the Electrolyte Potential at (a) Vg=-0.1 V (a) and at (b) Vg=+0.1 V after 60 sec. The rim of the gate electrode is outside the electrolyte and its potential is the biasing potential, Vg.

**Experiment and Methods:** Schematics of the wet-cell battery are shown in Fig. 3a-b. It measured 12 cm x 6 cm x 5 cm. It was made of two compartments: in one, a copper electrode was immersed in 0.1M $CuSO_4$. In the other, a zinc electrode was immersed in 0.1M $ZnSO_4$. The ion-bridge was replaced by a membrane (TS80 - polyamide filter made by Sterlitech) on which a thin film of 60:40, Au-Pd film was sputtered using a Hummer V sputtering system under Ar environment (this system is often used for SEM conductive film deposition). The resistance of the film was ca 25 KOhms/cm. The membrane was held tightly between two polymethyl methacrylate (PMMA) plates. One plate was glued to the cell while the other could be removed for gate membrane replacement. A 7-mm hole was drilled through each of the plates in order to let passage to the ions. A copper lead provided contact to the gate membrane. A simple paper filter, placed behind the gate membrane, enabled a better holding of the two plastic plates together.

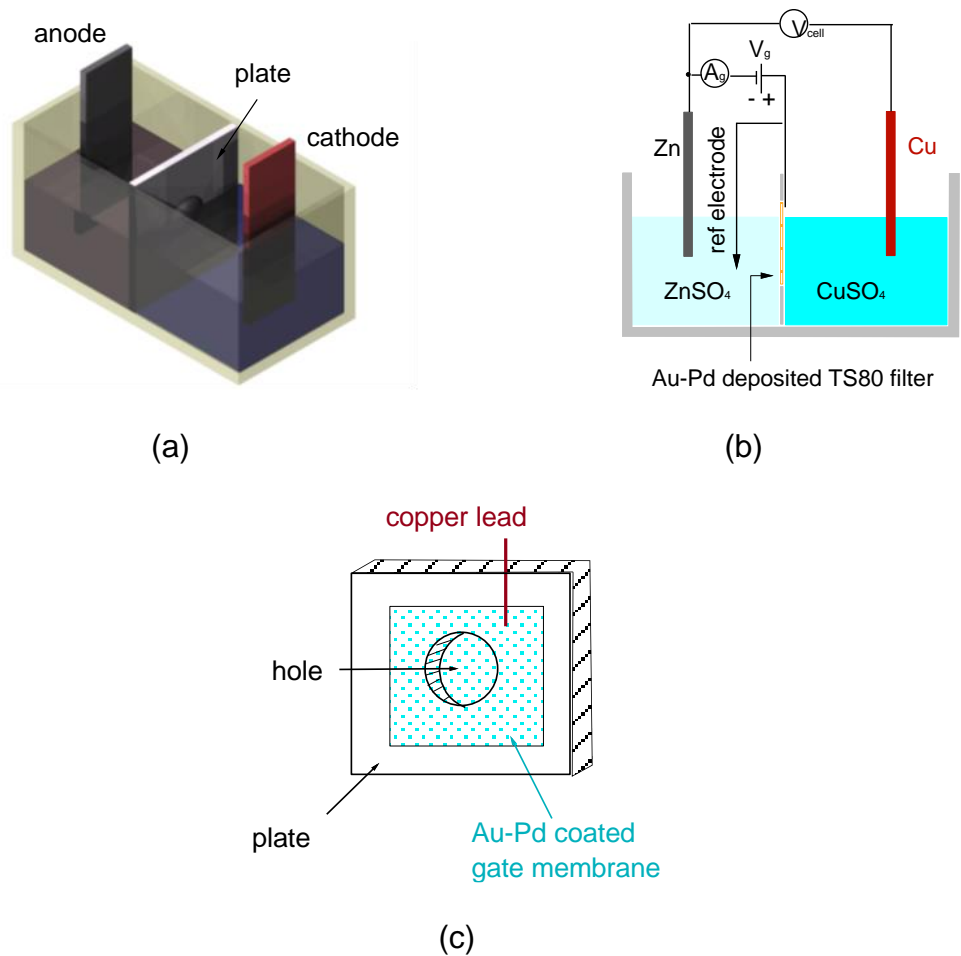

Fig. 3. (a) In our configuration the typical salt bridge is replaced by a permeable and conductive gate electrode (Au-Pd coated polyamide membrane). (b) The biased cell configuration. (c) A closer look at the removable plate, holder of the gate membrane.

**Cell's Characterization:** Characteristics of the cell are presented in Fig. 4. The open circuit cell's voltage was zero when the gate bias was $V_g \sim -0.3$ V (positive lead on the gate; negative/ground lead on the anode). At that point, the electronic gate current was $I_g \sim -1.3$ mA (namely, flowing from the gate to the cell's anode); the reference electrode exhibited ca -1.2 V between the gate and the $ZnSO_4$ electrolyte. The short circuit cell's current dropped to zero at $V_g \sim -0.3$ V, as well.

The copper contact wire to the gate, which was hidden behind the plastic plate was visually inspected after the conclusions of the experiments: it was clean and did not corrode after 11 days. Previous experiments also indicated that the type of material used for electrical contact between the gate's power supply and the gate electrode does not affect the stopping potential value provided that it is not exposed to the electrolyte [12]. Replacing the contact with a Co wire did not change the experimental results meaning that the contact material did not participate in any reaction at the gate.

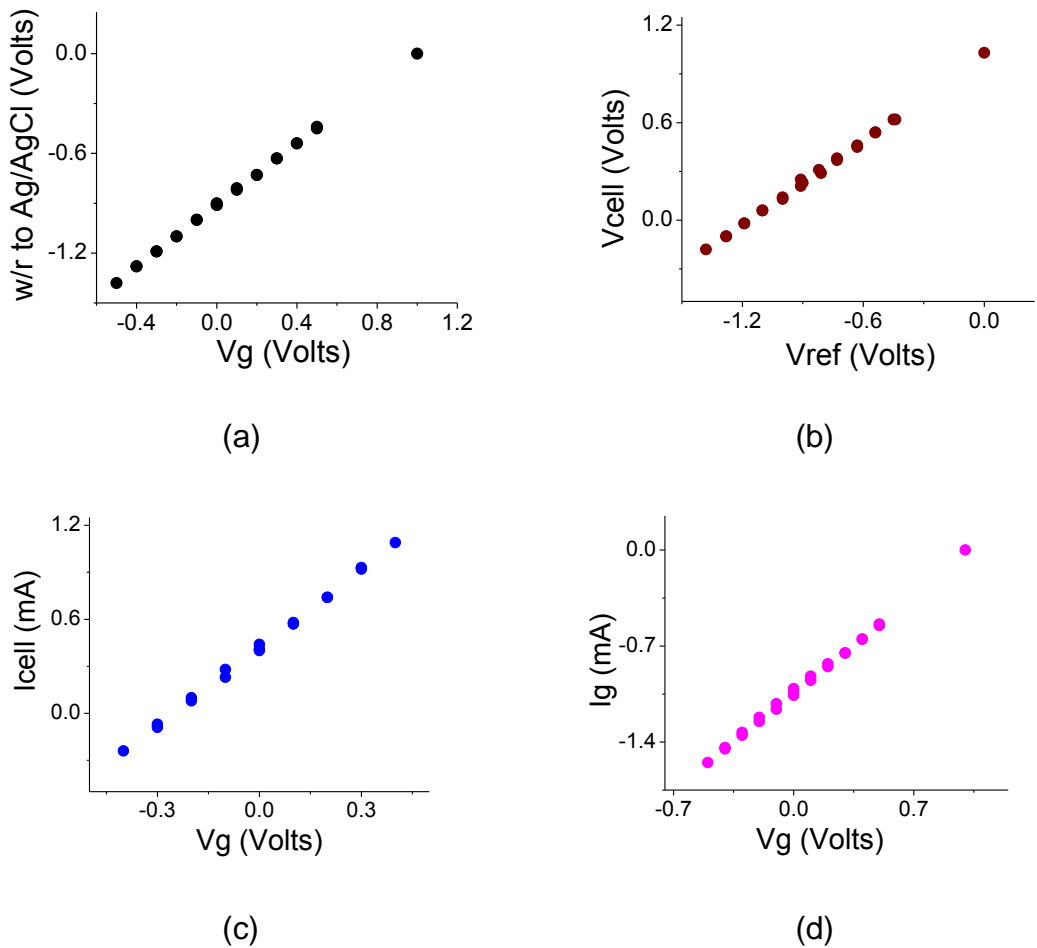

Fig. 4: (a) Calibrating the applied gate voltage with respect to an Ag/AgCl reference electrode. (b) Open circuit cell's voltage as a function of the gate voltage w/r to the reference electrode. (c) Short-circuit cell's current as a function of the gate voltage. (d) Gate current as a function of gate voltage. In all curves the scan was made in up and down directions (proving that no hysteresis is observed).

The gate electrode was biased with respect to the anode, which was also grounded. The gate membrane was tightly held between two plastic plates. A hole in each plate let the passage of ions. A copper wire, hidden behind the plates was providing a contact to the porous Au-Pd gate electrode (Fig. 1c). The copper wire remained clean and un-corroded after the conclusion of the experiments. Replacing the copper wire with a cobalt wire did not change the experimental results.

**The diode bridge:** Characteristics of a single and two Ge diodes connected in parallel, as well as, those of a Schottky diode are shown in Fig. 5. The 'break in' voltage, which can be assessed by the linearization of the large current region is important (and often ignored) in low voltage experiments. The choice of a diode should be such that its break-in forward voltage is below the desired signal. Schottky diodes, with a break-in forward voltage of 75 mV, were a good choice.

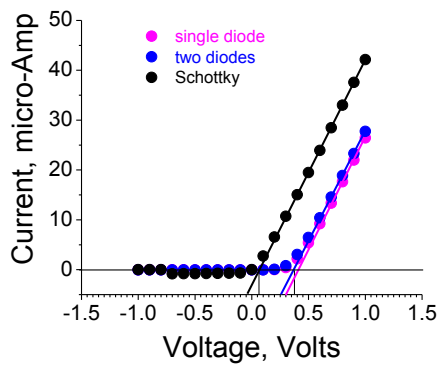

Fig. 5: Diodes' characteristics: a single Ge diode and two Ge diodes. 'Two diodes' refers to two Ge diodes connected in parallel in order to reduce the break-in voltage. The Schottky diode breaks in at $V_f$=75 mV.

**The Ion Gate:** The configuration for an ionic XOR gate is shown in Fig. 6a. The circuit was made of two anti-paralleled batteries, each incorporated with a gate electrode. If both ports, A and B were at ca 0 V, then their sum potential would be zero (see SI section). If one of the ports, say A was biased with 1 V, while the other was biased with 0 V, then the circuit output would be the difference between open circuit voltage of the 1 V biased battery and the 0 V biased battery  If both were biased with 1 V then the output of their sum potentials would be zero. The two 100 KOhms resistors were aimed at preventing current flowing directly from the cathode of one battery to the anode of the other. The diode bridge rectified any negative value occurring by the anti-parallel design. Our circuit is a bit different than the one reported earlier [11-12]: there, a graphene gate electrode was biased with respect to the cathode, and a positive bias of <+0.1 V sufficed to stop the battery from functioning. Here, an Au-PD electrode was biased with respect to the anode; a bias of ca -0.3 V stopped the battery from functioning. The effects are similar but of opposite Vcell-Vg trend. Finally, the cell's response is relatively fast, on the order of seconds.

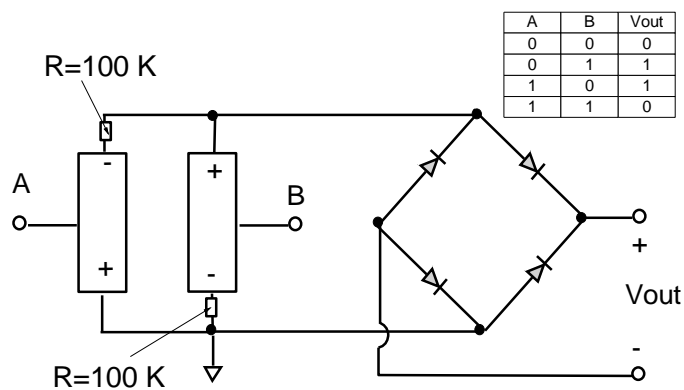

(a)

| A (V) | B (V) | Vout (V) |
|---|---|---|
| 0 | 0 | 0.006 |
| 0 | 1 | 0.7 |
| 1 | 0 | 0.7 |
| 1 | 1 | 0.006 |

(b)

| A | B | Vout |
|---|---|---|
| 0 | 0 | 0 |
| 0 | 1 | 1 |
| 1 | 0 | 1 |
| 1 | 1 | 1 |

(c)

| A (V) | B (V) | Vout (V) |
|---|---|---|
| 0 | 0 | 0.2 |
| 0 | 1 | 0.6 |
| 1 | 0 | 0.6 |
| 1 | 1 | 1.02 |

(d)

Fig. 6. (a) Anti-parallel battery configuration with a rectifying bridge makes an XOR ion logic gate. The inset shows the truth table for an XOR gate. (b) Experimental data for the circuit in (a). (c) The truth table for an OR gate. (d) Experimental data for the OR ion logic gate; the batteries were placed in-parallel and the resistors were removed.

**How these gates work?** The effect is attributed to capacitive and reactive processes at the gate electrode. At $V_{gate}=0$, the measured overpotential of the gate was ca -1.1 V whereas that of the cell's anode was higher, ca -0.9 V. This suggests that positive ions would flow from the gate to the cell's anode and electrical current would flow from the anode to the gate at zero biasing gate voltage, $V_{gate}=0$. Yet, the electrical gate current, $I_{gate}$ at $V_{gate}=0$ was ca -1.3 mA suggesting that the gate was serving as a cathode to the cell's anode (note that the positive side of the gate source was connected the gate and electrical current is defined by the flow of positive charges). We postulate that the energy invested at the gate is partially used to charge its capacitor and partially to counter the local reactions. The choice of a gate material with respect to the anode or the cathode is therefore important.

We can compare the power invested to stop the battery from functioning to the power supplied by the battery (the so called the Thevenin power, or $P_{TH}$). Specifically,

$$P_{TH}^{(cell)} = V_{cell}^{(OC)} \times I_{cell}^{(SC)} \text{ to be compared to } V_{gate} \times I_{gate}.$$

Here we define: the open-circuit voltage, $V_{cell}^{(OC)}$, the short-circuit current, $I_{cell}^{(SC)}$, and the related gate biasing voltage and current, $V_{gate}$ and $I_{gate}$. In our experiments: $P_{TH}^{(cell)} = V_{cell}^{(OC)} \times I_{cell}^{(SC)} = (1.05 \text{ V}) \times (1.23 \text{ mA}) = 1.29 \text{ mW}$ while the power invested to stop the battery from functioning was: $V_{gate} \times I_{gate} = (-0.3 \text{ V}) \times (-1.3 \text{ mA}) = 0.39 \text{ mW} \ll P_{TH}$. The battery was full-functioning again if the gate voltage was raised to +1.0 V; at that point, the gate current became zero (!) for zero invested power.

An OR gate is presented in Fig. 6c-d. We note that the OR gate has a reduction of ca 40% in the output voltage with only one functioning battery. Batteries are not pure voltage sources (whose voltage remain unchanged for all current levels consumed). Rather, the voltage of the battery changes depending on the current output [13]; in our case, by 40%.

**Summary:** electrical control over an electrochemical cell through mid-porous electrode enabled us to demonstrate self-powered Boolean logic: XOR and OR ion gates. This was achieved with an invested control power that was much smaller than the power produced by the battery itself.

**Acknowledgement:** To Jeff Urban from LBNL for invaluable comments.